\documentclass{article}
\newcommand{\bfr}{\begin{flushright}}
\newcommand{\efr}{\end{flushright}}
 
\begin{document}
\title{BORN-INFELD MONOPOLES AND INSTANTONS
}
\author{Atsushi Nakamula\\
Department of Physics, Tokyo Metropolitan University,\\
Minami-Ohsawa, Hachi-Oji-shi, Tokyo 192, Japan\\
and\\
Kiyoshi Shiraishi\\
Institute for Cosmic Ray Research, University of Tokyo\\
Midori-cho, Tanashi-shi, Tokyo 188, Japan
}
\date{HADRONIC JOURNAL {\bf 14} (1991) pp. 369--375
}
\maketitle
\begin{abstract}
We show that in certain non-Abelian Born-Infeld-Higgs theories there
are monopole solutions obeying Bogomol'nyi-like equations in the
Prasad-Sommerfield limit. We also discuss the existence of instanton
solution in an $SU(2)$ Born-Infeld theory.
\end{abstract}

Upon a spontaneous breaking of gauge symmetry containing the $SO(3)$
group, topologically stable configurations of gauge and scalar fields,
known as ``monopoles'' may form \cite{1}. The properties of the
nonsingular objects, such as their masses, charges, and spins, have
been closely investigated from the point of view of testing the grand
unineation scheme. Besides the phenomenological characteristics,
nathenatical methods for constructing the general monopole
configurations have also been studied with much interest. The most
interesting feature is the Bogomol'nyi equations \cite{2} in the
Prasad-Sommerfield limit \cite{3}. The study of the Bogomol'nyi
equations give not only the bound for monopole mass limit also the
proof of existence of multi-monopole solutions \cite{4}. The
interaction among monopoles is very simplified when the Bogomol'nyi
equations hold \cite{5}. The first-order equations is also intimately
related to study of supersymmetric systems \cite{6}.

Recently, the possibility that Yang-Mills interactions may receive
correction at high energy region has been argued along with studying
the string theory \cite{7}. As a candidate for the modification, a
generalized Born-Infeld (BI) action \cite{8} instead of Yang-Mills
action is preferred in effective field theories of open string theory
\cite{9}. The generalized-BI theory is expected to yield good
behavior of its classical solution and wave propagation similarly to
the original BI theory \cite{10}.

In our previous paper, we offered generalized (Abelian) Born-Infeld
models in which there are self-dual vortex solutions \cite{11}. In
the present paper, we search for monopole solutions in generalized BI
models. Our main problem is the determination of generalization of
Higgs fields. We postulate Bogomol'nyi-like equations from the
beginning; the Higgs potential is absent in our analyses, i.e., we
adopt the Prasad-Sommerfield limit \cite{3}. By this approach we
determine a non-minimal kinetic term of Higgs scalar.

First, we define a non-Abelian generalization of BI action. There is an
ambiguity in making an invariant form because of the ordering of
matrix-valued field strength. According to ref.~\cite{12}, a possible
form of Lagrangian is
\begin{equation}
L=\frac{\beta^2}{e^2}{\rm
Tr}\left[\exp\left\{-\frac{1}{2}{\rm
tr}\sum_{n=1}^\infty\frac{(-1)^n}{n}\left(\frac{\bf
F}{\beta}\right)^n\right\}-1\right]\,,
\end{equation}
where ${\rm Tr}$ is taken over the group indices while ${\rm tr}$ is
over the space-time indices. ${\bf F}$ represents matrix-valued
Yang-Mills field strength. $e$ is the gauge coupling constant while
$\beta$ is a parameter introduced here, which has dimension of
$({\rm mass})^2$. In this paper we take a simpler definition in four
dimensions;
\begin{equation}
L=\frac{\beta^2}{e^2}{\rm
Tr}\left\{\sqrt{1+\frac{1}{2\beta^2}{\bf F}_{\mu\nu}{\bf
F}^{\mu\nu}-\frac{1}{16\beta^4}({\bf F}_{\mu\nu}\tilde{\bf
F}^{\mu\nu})^2}-1\right\}\,.
\end{equation}
Here the square root should be recognized as Taylor series. As far as we
consider the configuration with spherical symmetry, an action which
reduces to the original BI action when the gauge symmetry is set to be
Abelian admits the same solution.

 Throughout this paper, we adopt an $SU(2)$ group as a non-Abelian
symmetry. Note that in the limit $\beta\rightarrow\infty$, the action
reduces to usual Yang-Mills action. We want to incorporate Higgs fields
into the action. It is natural to assume that {\it the model reduces to
the non-Abelian BI theory when the value of scalar fields are set to be
zero}. Further we do not take potentials for scalars into account;
i.e., we consider the Prasad-Sommerfield limit \cite{3}.

The Bogomol'nyi equations can be simply expressed as \cite{13}
\begin{equation}
B_i^a =(D_i\phi)^a\,,
\end{equation}
where $B_i^a=\frac{1}{2}F^{ajk}$ $(a=1, 2, 3)$ and $i, j, k=1, 2, 3)$
and
$D_i$ means the covariant derivative.

Since the monopole has purely magnetic
field, $(F\tilde{F})^2$ term is irrelevant for the solution in our BI
system. Then we find the following Lagrangian by inspection;
\begin{equation}
L=\frac{\beta^2}{e^2}{\rm
Tr}\left[\sqrt{1+\frac{1}{\beta^2}({\bf D}_\mu\phi)^2}
\sqrt{1+\frac{1}{\beta^2}({\bf F}_{\mu\nu})^2}-1
\right]\,.
\end{equation}

Strictly speaking, there are infinitely many possibilities in choosing
an action if we take the ordering of matrix-valued fields into account.
We consider here the simplest form of the action, keeping the
consistency with the usual $SU(2)$ Bogomol'nyi equations (3).

This action is reduced to a Yang-Mills action plus a canonical
kinetic term of a scalar field if we take the limit
$\beta\rightarrow\infty$. This model is a natural extension of
Yang-Mills-Higgs theory. The monopole solution is independent of the
value of $\beta$. It is notable that if all gauge fields vanish the
action of scalar fields has a good form for scalar wave propagation
\cite{14}.

It is easy to show the solution of the Bogomol'nyi equations saturates
the bound for the monopole mass. Since there is no electric field
around the monopole, the energy integral is written as
{\footnotesize
\begin{eqnarray}
E&=&\int d^3x\left[\frac{\beta^2}{e^2}{\rm
Tr}\left\{\sqrt{1+\frac{({\bf
D}_i\phi)^2}{\beta^2}}\sqrt{1+\frac{{\bf
B}^2}{\beta^2}}-1\right\}\right]\nonumber
\\ &=&\int d^3x\left[\frac{\beta^2}{2e^2}{\rm
Tr}\left\{\sqrt{1+\frac{({\bf
D}_i\phi)^2}{\beta^2}}\sqrt{1+\frac{{\bf
B}^2}{\beta^2}}\frac{1}{\beta^2}\left\{\frac{{\bf
B}_i}{\sqrt{1+\frac{{\bf
B}^2}{\beta^2}}}-\frac{{\bf
D}_i\phi}{\sqrt{1+\frac{({\bf
D}_i\phi)^2}{\beta^2}}}\right\}^2\right\}\right]\nonumber
\\ & &+\int d^3x\left[\frac{\beta^2}{2e^2}{\rm
Tr}\left\{\sqrt{1+\frac{({\bf
D}_i\phi)^2}{\beta^2}}\sqrt{1+\frac{{\bf
B}^2}{\beta^2}}\left\{\frac{1}{\sqrt{1+\frac{{\bf
B}^2}{\beta^2}}}-\frac{1}{\sqrt{1+\frac{({\bf
D}_i\phi)^2}{\beta^2}}}\right\}^2\right\}\right]\nonumber \\
& &+\int d^3x\left[\frac{1}{2e^2}{\rm Tr}({\bf B}\cdot{\bf
D}\phi+{\bf
D}\phi\cdot{\bf B})\right]\,.
\label{eq5}
\end{eqnarray}
}

Obviously we obtain the same mass as in the usual $SU(2)$ Yang-Mills
monopole with Prasad-Sommerfield limit \cite{13}, since the last term in
the right-hand side of eq.~(\ref{eq5}) gives the monopole mass so long
as the Bogomol'nyi equations hold.

Next we look for the Lagrangian which admits dyon solutions. In this
case $(F\tilde{F})^2$ term which has been ignored in the previous
analysis is essential.

The known equations are
\begin{eqnarray}
B_i^a&=&\cos \alpha\,(D_i\phi)^a\,,\\
E_i^a&=&\sin \alpha\,(D_i\phi)^a\,,
\label{eq7}
\end{eqnarray}
where $\alpha$ is a parameter which describes the ratio between
``magnetic'' and ``electric'' charges. In eq.~(\ref{eq7}), the left-hand
side means that $E_i^a=F_{i0}^a$. As a possible form of action to
satisfy the Bogomol'nyi equations, we find the following Lagrangian
for the dyon solution:
\begin{eqnarray}
L&=&\frac{\beta^2}{e^2}{\rm
Tr}\left[\left\{1+\frac{{\bf F}_{\mu\nu}{\bf
F}^{\mu\nu}}{2\beta^2}-\frac{({\bf
F}_{\mu\nu}\tilde{\bf
F}^{\mu\nu})^2}{16\beta^4}+\frac{({\bf
D}_\mu\phi)^2}{\beta^2}\right.\right.\nonumber \\
& &\qquad\qquad +\left.\left.
\frac{\tilde{\bf F}_{\mu}{}^{\nu}\tilde{\bf F}^{\mu}{}_{\lambda}{\bf
D}_\nu\phi{\bf D}_\lambda\phi}{\beta^4}
\right\}^{1/2}-1\right]\nonumber \\
&=&\frac{\beta^2}{e^2}{\rm
Tr}\left[\left\{1+\frac{{\bf B}^2-{\bf
E}^2+({\bf D}\phi)^2}{\beta^2}\right.\right.\nonumber \\
& &\qquad\qquad +\left.\left.
\frac{({\bf B}\cdot{\bf D}\phi)^2+({\bf E}\cdot{\bf D}\phi)^2
-({\bf
E}\cdot{\bf B})^2-{\bf E}^2({\bf D}\phi)^2}{\beta^4}
\right\}^{1/2}-1\right]\,.
\label{eq8}
\end{eqnarray}
This Lagrangian is also reduced to that of usual Yang-Mills-Higgs
Lagrangian when $\beta\rightarrow\infty$, while it reduces to the
non-Abelian BI action when the scalar fields vanish. Note that the
Lagrangian is not invariant under duality transformation, ${\bf
F}\leftrightarrow\tilde{\bf F}$, unless the scalar fields vanish.

We can find that the action coincides with the one obtained by
dimensional reduction of a five-dimensional non-Abelian BI action. The
Lagrangian (\ref{eq8}) can be derived from five-dimensional action
\begin{equation}
\int d^5x \left[\frac{\beta^2}{e^2}{\rm
Tr}\left\{\sqrt{-\det\left(\eta_{MN}+\frac{{\bf
F}_{MN}}{\beta}\right)}-1\right\}\right]\,,
\end{equation}
with ${\bf F}_{5i}={\bf D}_i\phi$.

This coincidence is not trivial because existence of topological
object depends crucially on dimensionality of system. Until now, we
have no convincing explanation to this fact. In the rest of this paper
we show the existence of instanton \cite{15} in the BI action.

Euclidean BI action is
\begin{equation}
S=\int d^4x\, {\rm
Tr} \,\frac{\beta^2}{e^2}\left(\sqrt{1+\frac{{\bf
B}^2+{\bf E}^2}{\beta^2}+\frac{({\bf B}\cdot{\bf
E})^2}{\beta^4}}-1\right)\,.
\label{eq10}
\end{equation}
Using notations
\begin{equation}
\bar{\bf D}=\frac{\partial S}{\partial{\bf E}}\quad\mbox{and}\quad
\bar{\bf H}=\frac{\partial S}{\partial{\bf B}}\,,
\label{eq11}
\end{equation}
we can express the equations of motion as
\begin{equation}
{\bf D}\cdot\bar{\bf D}=0\quad\mbox{and}\quad {\bf
D}\times\bar{\bf H}+\frac{\partial\bar{\bf D}}{\partial t}=0\,,
\label{eq12}
\end{equation}
where $t$ is the Euclidean time.

We impose (anti-) self duality in Yang-Mills field strength in
Euclidean space-time;
\begin{equation}
{\bf E}=\pm{\bf B}\,.
\end{equation}
Then according to action (\ref{eq10}), we obtain
\begin{equation}
\bar{\bf D}={\bf E}=\pm{\bf B}\quad\mbox{and}\quad
\bar{\bf H}={\bf B}=\pm{\bf E}\,.
\end{equation}
Thus the equations of motion (\ref{eq12}) reduces to the Bianchi
identity
\begin{equation}
{\bf D}\cdot{\bf B}=0\quad\mbox{and}\quad
{\bf D}\times{\bf E}+\frac{\partial{\bf B}}{\partial t}=0\,.
\end{equation}
From this result, we conclude that usual Yang-Mills instanton solution
\cite{15} is still a solution in Euclidean non-Abelian BI theory.

In future work, we wish to consider supersymmetric generalization of
non-Abelian BI-Higgs system. It will also be interesting to analyze
scattering of femions by topological defects in such a system.
Construction of multi-monopole configuration may have subtleties
because of the non-polynomial nature of the action in our case.

Inclusion of gravity may be highly non-trivial and the functional
form of monopole solution will be distinct from that of usual
Yang-Mills monopole.

We wonder whether effective field theory derived from four-dimensional
open string theory \cite{16} is governed by BI-Higgs action. If not, the
action analyzed in the present paper may be a starting point to analyze
the deviation from it.

\section*{Acknowledgments}
The authors would like to thank S. Hirenzaki for some useful
comments. One of the authors (KS) would like to thank A. Sugamoto
for reading this manuscript.

KS is indebted to Soryuusi shogakukai for financial support. He also
would like to acknowledge financial aid of Iwanami F\=ujukai.


\end{document}